\documentclass[aip,jmp,onecolumn,reprint,leqn]{revtex4-2}

\usepackage{amsmath,amsfonts,amssymb,mathrsfs,graphicx,mathtools,dcolumn,bm}
\usepackage[colorlinks,linkcolor=blue,citecolor=blue,urlcolor=blue]{hyperref}
\usepackage[usenames,dvipsnames]{color}

\usepackage[utf8]{inputenc}
\usepackage[T1]{fontenc}
\usepackage{etoolbox}

\newcommand{\RNum}[1]{\uppercase\expandafter{\romannumeral #1\relax}}

\renewcommand\a{\alpha}

\renewcommand\gg{\gamma}
\newcommand\del{\delta}

\newcommand\e{\eta}
\newcommand\q{\theta}

\renewcommand\l{\lambda}

\newcommand\ff{\varphi}
\renewcommand\r{\rho}

\newcommand\f{\phi}
\renewcommand\c{\chi}
\renewcommand\j{\psi}

\renewcommand\L{\Lambda}
\newcommand\J{\Psi}

\newcommand{\oo}{\mathscr{O}}
\newcommand{\ry}{\mathcal{R}}
\newcommand{\hh}{\mathcal{H}}
\newcommand{\ho}{\mathcal{H}_{\oo}}
\newcommand{\no}{\mathcal{B}_{\oo}}
\newcommand{\ro}{\mathcal{R}_{\oo}}

\def\rr{\mathbb{R}}
\def\cc{\mathbb{C}}

\renewcommand\d{\partial}
\newcommand\ra{\rightarrow}

\newcommand{\nmb}{\nonumber}
\newcommand\bea{\begin{align}}
\newcommand\eea{\end{align}}

\makeatletter
\def\@email#1#2{%
	\endgroup
	\patchcmd{\titleblock@produce}
	{\frontmatter@RRAPformat}
	{\frontmatter@RRAPformat{\produce@RRAP{*#1\href{mailto:#2}{#2}}}\frontmatter@RRAPformat}
	{}{}
}%
\makeatother

\begin{document}
	
	\title{Generalised Geometric Phase: Mathematical Aspects}
	
	\author{Vivek M. Vyas}
	\email{vivek.vyas@iiitv.ac.in}
	\affiliation{Indian Institute of Information Technology Vadodara, Government Engineering College, Sector 28, Gandhinagar 382028, India}

\begin{abstract}
An operator generalisation of the notion of geometric phase has been recently proposed purely based on physical grounds. Here we provide a mathematical foundation for its existence, while uncovering new geometrical structures in quantum systems. While probing the average of any observable it is found that a quantum system exhibits different ray spaces and associated fibre bundle structures. The generalised geometric phase is understood  as (an)holonomy of a connection over these fibre bundles. The underlying ray spaces in general are found to be pseudo-K\"ahler manifolds, and its symplectic structure gets manifests as the generalised geometric phase.
\end{abstract}


\date{\today}
\maketitle

\section{Introduction}

The last century has witnessed intense activity on the quest of finding the underlying mathematical structure of quantum physics \cite{duck2000,ecgbook}. From the pioneering work of Dirac, it became clear that the abstract Hilbert space provides the foundation to all the erstwhile known formulations of quantum mechanics \cite{diracbook, van2007}. This realisation lead to a rapid growth in the development of quantum physics, wherein the mathematical framework of vector space and linear operators was principally employed.

One of key ideas in this framework is that the physical state of any given quantum system is represented by a corresponding vector $\J$ in the Hilbert space $\hh$ of the system. Soon it was realised that this correspondence is not one-to-one, and not all the vectors in $\hh$ represent distinct physical states of the system. In particular any two collinear vectors: $\J$ and $c \J$ ($c \in \cc$) describe the same physical state. Thus it was realised that while the Hilbert space structure of quantum physics has provided with deep insights, there exists an infinite redundancy and ambiguity in the existing mathematical framework. A straightforward way to get rid of this redundancy is by identifying all the collinear vectors as one entity, generally referred as \emph{rays}. Mathematically this identification essentially defines an equivalence class $\ry = \hh/\sim$ where $\J \sim c \J$. Owing this identification the set $\ry$ losses the vector space structure, and the objects that form $\ry$ are no longer vectors in $\hh$. Rather they are linear operators over $\hh$, popularly known as the \emph{pure state density matrices} and defined as $\r = \frac{\J \J^{\dagger}}{(\J, \J)}$. It is interesting to note that historically the notion of density matrix was necessitated in order to formulate physical problems pertaining to the realm of open systems, and systems at thermal equilibrium \cite{fano1957}.

Knowing the fact that the space $\ry$ of pure state density matrices does not form a linear vector space, eventually led to the natural question so as to what is the mathematical structure of the $\ry$ ? In a landmark work, Kibble showed that $\ry$ is a complex manifold with the existence of non-degenerate and closed two-form giving rise to a symplectic structure akin to the phase space of classical Hamiltonian mechanics \cite{kibble1979}. In another major development, 
Provost and Vallee showed that $\ry$ is also admits a Riemannian metric structure coexisting with the symplectic structure \cite{provost1980}. In some cases it was noted that the Riemannian metric  over $\ry$ is related to the uncertainty or variance of certain observables, which clarified its physical significance.  However the physical meaning and the manifestation of the symplectic structure was yet to be understood.

In 1984, Berry's celebrated work on what is now known as the \emph{geometric phase} caught the attention of the scientific community\cite{berry1984,bsimon1983,anandan1987,anandan1988geometric}. As noted by Berry, the concept of geometric phase had been anticipated independently by several preceding workers \cite{berry1990,shapere1989}. In order to appreciate the mathematical relevance of the geometric phase in its simplicity, let us consider a quantum system that admits a closed continuous curve $C$ of unit vectors $\j(s)$ in the corresponding Hilbert space $\hh$ parametrised by real number $0\leq s \leq L$ such that $\j(0) \equiv \j(L)$.  The geometric phase corresponding to such a curve is then defined by:
\begin{align*}
	\gamma = \oint_{C} \: ds A(s), 
\end{align*}
where $A(s) = \text{Im} (\j(s), \f(s))$, where $\f(s) = \frac{d \j(s)}{ds}$ \cite{berry1984,mukunda1993,samuel1}.  While this closed integral is defined using unit vectors, eventually it became clear that it can be brought down to the corresponding ray space, and can be written as a surface integral of the symplectic two-form $\omega$:
\begin{align*}
	\gamma = \int_{\mathcal{S}} \omega.
\end{align*}
Here $\mathcal{S}$ represents a surface in the ray space, which is bounded by the image of $C$. Thus it was found that the geometric phase essentially captures the symplectic structure of the ray space \cite{anandan1988geometric,mukunda1993,page1987,arnobohm}. For a circuit that is formed by connecting any given three rays $\r_1$, $\r_2$ and $\r_3$ by geodesics, the geometric phase can be neatly expressed as\cite{mukunda1993,mukunda2003null}:
\begin{align*}
	\gg = \text{Arg} \left( \r_1 \r_2 \r_3\right).
\end{align*}

The geometric phase has been experimentally measured and studied in various contexts \cite{agarwal1990,anandan1992,berry1987,chiao1986,samuel2,mathur1991,shapere1989}. Most notably the current understanding of the topological materials crucially relies on its existence \cite{bernevig2013,bsimon1983,thouless1985,thouless1982,thouless1983}.   
We now understand that the ray space $\ry$ and the vector space $\hh$ together form what is called the principal fibre bundle structure, whose manifestation is seen in the geometric phase treatment, wherein $A(s)$ is understood as a connection over the fibre bundle, and $\gg$ as its (an)holonomy \cite{arnobohm,anandan1988geometric,bertlmann2000anomalies}. 

In a recent work, an operator generalisation of the concept of geometric phase for pure states was proposed by the author, and its physical manifestations were shown \cite{vyas2023}. This generalisation essentially arose from the fact that the interference phenomenon also manifests in the average of the observables, which is captured by the argument of the matrix elements of an observable. This gives rise to identification of this argument as a generalised relative phase.  Following the trail of Pancharatnam, from generalised relative phase one is naturally lead to the generalised notion of geometric phase \cite{mukunda1993}. It was explicitly shown that this generalised geometric phase can solely be expressed in terms of density matrices. For any given three density matrices $\r_1$, $\r_2$ and $\r_3$, the generalised geometric phase corresponding to an observable $\oo$ is given by:
\begin{align} \label{ggp3}
	\gg_{\oo} = \text{Arg} \: \text{Tr}\left( \frac{\r_1 \oo \r_2 \oo \r_3 \oo}{\langle \oo \rangle_{1} \langle \oo \rangle_{2} \langle \oo \rangle_{3}} \right),
\end{align}
where $\langle \oo \rangle_{j} = \text{Tr} (\r_{j} \oo) \neq 0$ ($j=1,2,3$). For a closed curve $C$ as defined earlier, the generalised geometric phase $\gamma_{\oo}$ corresponding to an observable $\oo$ is:
\begin{align*}
	\gamma_{\oo} = \oint_{C} \: ds A^{\oo}(s), 
\end{align*}
where $A^{\oo}(s) = \text{Im} \frac{ (\j(s), \oo \f(s))}{(\j(s),\oo \j(s))}$. This geometric phase is well defined so long as $(\j(s),\oo \j(s)) \neq 0$.  It can be easily seen to be invariant under continuous local gauge transformation $\j(s) \rightarrow e^{i \Lambda(s)}\j(s)$, for any smooth function $\Lambda(s)$ which is periodic $\Lambda(0) = \Lambda(L)$. This clearly indicates that it is a well defined geometric object over the ray space. As discussed above, the geometric phase $\gamma$ essentially captures the symplectic structure of the underlying ray space $\ry$ of the given quantum system. Then one wonders what geometrical aspect of the ray space is captured by $\gamma_{\oo}$.

The main goal of this paper is to address and answer this question. We show that the operator generalised geometric phase exists due to the presence of ray spaces other than $\ry$, and it captures the symplectic structure defined over them. The existence of these ray spaces $\ro^{\pm}$ is understood by noting that $\hh$ can be divided into three disjoint sets: a) vectors with $(\j, \oo \j) > 0$,
b) vectors with $(\j, \oo \j) < 0$, and c) vectors with $(\j, \oo \j) = 0$. The vectors in the first two sets can be normalised so that all the vectors respect the condition $(\j, \oo \j) = \text{sgn} (\j, \oo \j)$. This paves the way for the identification of collinear vectors in the respective sets, to yield the ray spaces $\ro^{\pm}$. This gives us two independent principal fibre bundle structures, which are in general different than the one with $\ry$. It is found that the object $A^{\oo}$ is a connection over these new fibre bundles. Akin to $\ry$, the ray spaces $\ro^{\pm}$ are found to possess a symplectic structure, and also have a co-existing pseudo-Riemannian metric, which makes them pseudo-K\"ahler manifolds. It is seen that essentially the generalised geometric phase $\gamma_{\oo}$ is the (an)holonomy of the connection $A^{\oo}$, which captures the symplectic structure of the ray spaces $\ro^{\pm}$.    

In the next section, a short discussion on the mathematical framework behind the existence and the manifestation of the usual geometric phase is presented. In section (\ref{ggp}), the mathematical framework responsible for the existence of the operator generalised geometric phase is discussed. The paper concludes with a short summary of the present work. 

\section{Geometry of the Geometric Phase \label{gprev}}

In this section, for the benefit of the readers, we shall briefly review the mathematical framework that is responsible for the genesis of the geometric phase. A detailed mathematical treatment of this subject appears in Refs. \onlinecite{mukunda1993,samuel1,page1987,anandan1988geometric,arnobohm}.

\subsection{Fibre bundle structure}
Consider that we are given a quantum system with the Hilbert space $\mathcal{H}$. We shall generally assume that the Hilbert space is separable, with the complex dimension $N$, which can be finite or infinite. Let the collection of unit vectors be denoted as $\mathcal{B}$:
\begin{align}
    \mathcal{B} = \{ \j \in \mathcal{H} \lvert (\psi, \psi) = 1 \}.
\end{align}
This subset of $\mathcal{H}$ does not form a vector space. Any non-zero vector in $\mathcal{H}$ can be unit normalised and brought down to this subset. 

Expanding a general state $\psi$ in an orthonormal basis $\{ {e}_{i} \}$ (for $i=1,2,\cdots,N$) we get $\j = \sum_{i=1}^{N} x_{i} {e}_i $, where the complex coefficients are $x_i = ({e}_i,\j)$. The complex coefficients $\{ x_{i} \}$ depicting any unit vector in $\mathcal{B}$ are constrained to obey the algebraic equation $\sum_{i=1}^{N} |x_i|^2 = 1$. This shows that the set $\mathcal{B}$ is a Euclidean sphere $S^{2N-1}$ of real dimension $(2N-1)$, as also a smooth simply connected manifold. 

While mathematically any two unit vectors $\f$ and $e^{i \q} \f$ belonging to $\mathcal{B}$ are distinct for real $\q \neq 0$, they are \emph{physically indistinguishable}. As a result one defines the space of rays $\mathcal{R}$ associated with $\mathcal{B}$, wherein any two vectors in $\mathcal{B}$ differing by a $U(1)$ phase are identified. Thus ray space $\mathcal{R}$ is a quotient of $\mathcal{B}$ with respect to the equivalence relation $\f \sim e^{i \q} \f$ (for $\q \in \rr$): $\mathcal{R} = \mathcal{B}/\sim$. This space is often referred to as a projective Hilbert space. The elements in $\mathcal{R}$ called \emph{rays}, and are specified by the pure state density matrices $\r_{\j} = \j \j^{\dagger}$. This defines a projection $\pi$ from $\mathcal{B}$ to $\mathcal{R}$:
\begin{align} \nmb
    \pi(\j) = \r_{\j} = \j \j^{\dagger} \in \mathcal{R} \: \text{for all} \: \j \in \mathcal{B}.
\end{align}
The triple $(\mathcal{B},\mathcal{R},\pi)$ constitutes a principle fibre bundle over the total space as $\mathcal{B}$, base space as $\mathcal{R}$ and with $U(1)$ as the structure group \cite{mukunda1993,samuel1,bertlmann2000anomalies,anandan1988geometric,bsimon1983}. A fibre consists of all unit vectors belonging to the same ray. This mathematical structure is of great importance in quantum mechanics, and forms the basis for the existence of the geometric phase\cite{anandan1988geometric,samuel1,shapere1989,bsimon1983}.

Consider a smooth curve $\J(l)$ in $\mathcal{B}$ which is parametrised by $l$, and passing through a particular point specified by $\j$, so that at $l=L$(say): $\J(L) \equiv \j$. Since $(\J(l),\J(l)) = 1$, we have that $\text{Re} (\J(L),\dot{\J}(L)) = 0$. This motivates one to define the tangent space $T_{\j}\mathcal{B}$ at any point $\j$ on the sphere $\mathcal{B}$ as:
\begin{align} \nmb
    T_{\j}\mathcal{B} = \{ \f \in \mathcal{H} \lvert \text{Re} (\j, \f) = 0  \}.
\end{align}
It can be seen that $T_{\j}\mathcal{B}$ is actually a real linear vector space of dimension $(2N-1)$. If a tangent vector to $\j$ is also collinear to it, then it is said to belong to the vertical subspace $V_{\j}\mathcal{B} \subset T_{\j}\mathcal{B}$, which is defined as:
\begin{align} \nmb
	V_{\j} \mathcal{B} = \{ i a \j \lvert \: a \in \rr \}.
\end{align}
This sets the ground to define the \emph{connection} $A_{\j}(\f)$ on $\mathcal{B}$ at point $\j$ as a linear functional on the tangent space $T_{\j}\mathcal{B}$, given by:
\begin{align} \label{connecdef}
    A_{\j}(\f) = \text{Im} (\j,\f), \: \text{where} \: \f \in T_{\j}\mathcal{B}. 
\end{align}
The horizontal subspace $H_{\j}\mathcal{B} \subset T_{\j}\mathcal{B}$ consists of those tangent vectors for which the connection vanishes:
\begin{align} \nmb
  H_{\j}\mathcal{B} = \{ \f \in T_{\j}\mathcal{B} \lvert \: A_{\j}(\f) = 0 \}.  
\end{align}
This is a linear vector space of real dimension $2(N-1)$. It can be seen that any tangent vector can be uniquely decomposed into components belonging to the vertical subspace and horizontal subspaces so that one has $T_{\j} \mathcal{B} = V_{\j}\mathcal{B} \oplus H_{\j}\mathcal{B}$.

Consider an open neighbourhood $\mathcal{M}$ in $\mathcal{R}$ defined around point $\r_0 = \j_0 \j^{\dagger}_{0}$, such that $\mathcal{M} = \{ \r_{\j} = \j \j^{\dagger} \in \mathcal{R} \: \lvert \: \text{Tr } \r_{0} \r_{\j} > 0 \}$. Any point $\r_\j$ in this neighbourhood can be specified through:
\begin{align} \label{psivector}
    \j(\a,\xi_0) = e^{i \a} \tilde{\j} = e^{i \a} \left( \sqrt{1 - (\xi_0, \xi_0)} \j_0 + \xi_0 \right),
\end{align}
where $\a$ is real, and $\xi_0 \in H_{\j_0}\mathcal{B}$ such that $(\xi_0, \xi_0) < 1$. The set of vectors $\j(\a,\xi_0)$ indeed also define an open neighbourhood in $\mathcal{B}$, which can be referred to as $\pi^{-1}(\mathcal{M})$, which is coordinatised by $\a$ and $\xi_0$. It is evident that all the vectors in $\pi^{-1}(\mathcal{M})$ project onto $\mathcal{M}$.      

When $\a$ and $\xi_0$ are varied infinitesimally the vector $\j(\a,\xi_0)$ changes as:
\begin{align} \nmb
\del \j = i \del \a \j + e^{i \a} \left( \del \xi_0 - \j_{0} \frac{(\xi_0, \del \xi_0)}{\sqrt{1 - (\xi_0, \xi_0)}}\right). 
\end{align} 
Suppose if this change in $\j$ is solely along some tangent vector $\f = i a \j + \xi$ (here $a \in \rr$ and $(\xi, \j) = 0$), so that $\del \j = \varepsilon \f $, where $\varepsilon$ is an infinitesimal. This determines the variations $\del \a$ and $\del \xi_0$ as:
\begin{align} \nmb
	\del \a &= \varepsilon \left( a - \frac{\text{Im} (\xi_0, \tilde{\xi})}{1 - (\xi_0, \xi_0)}   \right),\\ \nmb
	\del \xi_0 &= \varepsilon \left( \tilde{\xi} + i \tilde{\j} \frac{\text{Im} (\xi_0, \tilde{\xi})}{1 - (\xi_0, \xi_0)} + \j_{0} \frac{\text{Re} (\xi_0, \tilde{\xi})}{\sqrt{1 - (\xi_0, \xi_0)}}\right).
\end{align}
Here $\tilde{\xi} = e^{-i \a} \xi$ and $\tilde{\j} = e^{-i \a} \j$. Using these two relations one can setup a differential geometric representation of tangent vector $\f$ as:
\begin{align} \nmb
	X_{\f} = \frac{\del \a}{\varepsilon} \frac{\d}{\d \a} + \frac{\del \xi^{\dagger}_0}{\varepsilon} \frac{\d }{\d \xi^{\dagger}_0} + \frac{\d }{\d \xi_0} \frac{\del \xi_0}{\varepsilon}.
\end{align}
The connection $A_{\j}(\f)$ can be represented as a \emph{one-form $A$} at point $\j(\a,\xi_0)$ as:
\begin{align} \label{forma}
    A = d\a - \frac{i}{2} \left( \xi_{0}^{\dagger} d \xi_{0} -  d \xi_{0}^{\dagger} \xi_{0} \right),
\end{align}
so that its interior product with $X_{\f}$ gives $A_{\j}(\f)$:
\begin{align} \nmb
	i_{X_{\f}} A = A_{\j}(\f) = \text{Im} (\j,\f).
\end{align}
Owing to the appearance of $d \a$ term in expression (\ref{forma}) it can be seen that one-form $A$ is defined over $\mathcal{B}$ and can not expressed as a pullback via $\pi^\ast$ for any one-form defined over $\mathcal{R}$. In other words, $A$ can not be defined over the ray space.

Suppose one is given a closed curve $C$ in the neighbourhood $\pi^{-1}(\mathcal{M})$ specified as:
\begin{align} \nmb
    C = \{ \j(s) \in  \pi^{-1}(\mathcal{M}) \lvert \: 0 \leq s \leq \L, \: \text{and}\:\j(0) = \j(\L)\}.
\end{align}
Then the \emph{geometric phase} $\gg[C]$ corresponding to such a curve is defined as the (an)holonomy of the connection along the curve \cite{mukunda1993,samuel1,mukunda2003null,rabei1999}:
\begin{align} \nmb
    \gg[C] &= \oint_{C} \: ds \: A_{\j(s)}(\dot{\j}(s)) \\ \nmb &=\oint_{C} \: ds \: \text{Im} (\j(s),\dot{\j}(s)) \\
           &= \oint_{C} \: A, 
\end{align}
where $A$ is given by (\ref{forma}). By the virtue of Stokes theorem this loop integral can be traded off in favour of a surface integral of the two-form $dA$ over a smooth surface $S$ in $\mathcal{B}$ whose boundary is $C$:
\begin{align} \nmb
    \gg[C] &= \int_{S} \: dA.
\end{align}
From (\ref{forma}) it immediately follows that the two form $dA$ is independent of $d\a$, and it reads: 
\begin{align} \label{twoformda}
	dA = - i d \xi_{0}^{\dagger} \wedge d\xi_{0}.
\end{align}
Consider two tangent vectors $\f_{i}$ (for $i=a,b$) at point $\j$, which are given by: $\f_{i} = i \a_{i} \j + \xi_{i}$ (where $\a_{i} \in \rr$ and $(\xi_{i},\j)=0$). One can then construct their corresponding differential geometric representations $X_{\f_{i}}$ following the treatment presented earlier. The two form $dA$ can be expressed as a bilinear antisymmetric functional:
\begin{align} \nmb
	dA_{\j} (X_{\f_{a}},X_{\f_{b}}) &= i_{X_{\f_{a}}} i_{X_{\f_{b}}} dA = 2 \: \text{Im} (\f_{a},\f_{b}) \\ &= 2 \: \text{Im} (\xi_{a},\xi_{b}).   \label{twoformfunc}
\end{align} 
Since $dA$ is not dependent on $\a$ so it can be expressed as a pullback of some two-form $\omega_{\r}$ defined on $\mathcal{R}$:   $dA = \pi^{\ast} \omega_{\r}$. This allows the geometric phase to be expressible as a surface integral over $\pi(S)$, which is the image of $S$ in $\mathcal{R}$: 
\begin{align}
	\gg[C] = \int_{S} \: dA = \int_{\pi(S)} \omega_{\r}.
\end{align}
Noting that the density matrix corresponding to any vector $\j(\a,\xi_{0})$ is solely a function of horizontal vectors $\xi_{0}$ and $\xi^{\dagger}_{0}$, one sees that $\xi_{0}$ and $\xi^{\dagger}_{0}$ actually coordinatize the neighbourhood $\mathcal{M}$ in the ray space. One thus identifies the two-form as $\omega_{\r} = - i d \xi_{0}^{\dagger} \wedge d\xi_{0}$. The two form $\omega_{\r}$ is closed but not exact.

One might wonder if it is possible to express the two-form $\omega_{\r}$ as a bilinear antisymmetric functional solely in terms of intrinsic ray space objects. To do so one first needs to define the tangent space $T_{\r}\mathcal{R}$ corresponding to point $\r_\j$ in the ray space $\mathcal{R}$. It is worth recalling that the pure state density matrices $\r_{\j} = \j \j^\dagger$ forming the ray space are such that: $\text{Tr} \: \r_{\j} = 1 \: \& \: \r^{2}_{\j} = \r_{\j}$.

Consider a smooth curve $\ff(l)$ in $\mathcal{B}$, which is parametrised by $l$, and for which one has a smooth curve in $\mathcal{R}$ given by $\r_{\ff} = \ff(l) \ff^{\dagger}(l)$. At any particular point $l=q$, the tangent vector $B_{\ff}(q) = \dot{\r}_{\ff}(q)$ is an operator such that $\text{Tr} \: B_{\ff}(q) = 0$, and is explicitly given by $B_{\ff}(q) = \dot{\ff}(q) \ff^{\dagger}(q) + \ff(q) \dot{\ff}^{\dagger}(q)$. This allows one to define the tangent space $T_{\r}\mathcal{R}$ at any point $\r$ explicitly in terms of $H_{\j}\mathcal{B}$ as:
\begin{align} \label{raytangent}
    T_{\r}\mathcal{R} = \{ B = \xi \j^{\dagger} + \j \xi^{\dagger} \: \lvert \: \xi \in H_{\j}\mathcal{B} \}.
\end{align}
In terms of two elements $B'$ and $B''$ belonging to $T_{\r}\mathcal{R}$, which correspond respectively to horizontal vectors $\xi'$ and $\xi''$, the two-form $\omega_\r$ at point $\r$ is specified as an antisymmetric bilinear functional: 
\begin{align} \label{twoformro}
    \omega_{\r}(B',B'') &= -i \text{Tr} (\r [B',B''])\\ &= 2 \; \text{Im} (\xi',\xi'').
\end{align}
This shows that while both $A$ and $dA$ are globally defined forms on $\mathcal{B}$, only $dA$ is projectable to $\mathcal{R}$. 

Interestingly the two form $\omega_{\r}$ defines a \emph{symplectic structure} on the ray space $\mathcal{R}$. This can be easily seen by considering any orthonormal basis $\{ \tilde{e}_{j} \}$ which span $H_{\j_0} \mathcal{B}$, which is the orthogonal complement of $\j_{0}$. Expanding the state $\xi_{0}$ in this basis we get:
\begin{align} \nmb
	\xi_{0} = \sum_{j} \frac{1}{\sqrt{2}}(q_{j} - i p_{j}) \tilde{e}_{j}, 
\end{align} 
where the real coefficients $q_{j}$ and $p_{j}$ provide a local coordinate system over the neighbourhood $\mathcal{M}$ of the ray space. In terms of these real coordinates, the two-form takes the canonical form:
\begin{align} \label{twoform}
	\omega_\r = \sum_{j} dq_{j} \wedge dp_{j},
\end{align}
explicitly displaying its non-degeneracy.

This shows that the genesis of the geometric phase lies in the existence of the symplectic structure on the ray space.

Noting the above symplectic structure on the ray space $\mathcal{R}$, one might ask if there is any other geometrical structure on $\mathcal{R}$. It turns out that the ray space also admits a Riemannian metric structure. To appreciate this let us consider some smooth curve in $\mathcal{B}$ specified by vector $\ff(s)$ and parametrised by $s$. The difference between the two vectors $\ff(s)$ and $\ff(s + ds)$  due to infinitesimal change $ds$ is then given by $\ff(s + ds) - \ff(s) = \left( \frac{d \ff}{ds} \right) ds$. The squared norm of the difference $(\ff(s + ds) - \ff(s))$ gives a notion of distance $dl^2$ between the two states, which can be written as:
\begin{align} \nmb
	dl^2 &= (\ff(s + ds) - \ff(s), \ff(s + ds) - \ff(s)) \\ \nmb
	&= \left(\frac{d \ff}{ds}, \frac{d \ff}{ds} \right) ds^2. 
\end{align}
This distance however is not projectable to the ray space, and can not be interpreted as a distance between $\pi(\ff(s))$ and $\pi(\ff(s + ds))$. This is owing to the fact that under a phase transformation $\ff(s) \rightarrow \ff'(s) = e^{i \Lambda(s)} \ff(s)$, the projection is invariant $\pi(\ff(s)) \equiv \pi(\ff'(s))$, while the distance $dl^2$ is not. In order to circumvent this problem one employs the covariant derivative of $\ff$ defined as:
\begin{align}
	\frac{D \ff(s)}{ds} = \frac{d \ff(s)}{ds} - i A_{\ff}(s) \ff(s),
\end{align}
 where $A_{\ff}(s) = \text{Im} (\ff, \frac{d \ff}{ds})$\cite{samuel1}. It can be readily checked that this covariant derivative lies in the horizontal subspace $H_{\ff}\mathcal{B}$ since $(\ff(s), \frac{D \ff(s)}{ds}) = 0$. One can now define a notion of distance $d\mathcal{L}^2$ between $\pi(\ff(s))$ and $\pi(\ff(s + ds))$ as:
\begin{align} \label{dist}
	d\mathcal{L}^{2}(\pi(\ff(s)),\pi(\ff(s + ds))) = \left(\frac{D \ff}{ds}, \frac{D \ff}{ds} \right) ds^2.
\end{align}
Thus we have a Riemannian metric $g(\xi',\xi'')$ on $\mathcal{R}$, which is a positive semidefinite symmetric bilinear functional of any two horizontal vectors $\xi',\xi'' \in H_{\j}\mathcal{B}$ at point $\j \in \mathcal{B}$, given by:
\begin{align}
	g_{\r_\j}(\xi',\xi'') = \text{Re} (\xi',\xi''). 
\end{align}
This metric can be expressed solely in terms of ray space objects by noting that the tangent vector $B $ to any point $\r_{\j} \in \mathcal{R}$ is expressible in terms of horizontal vector $\xi$ via the relation $B = \xi^{\dagger} \j + \j^{\dagger} \xi$. It is straightforward to see that at point $\r_{\j}$:
\begin{align} \label{metric}
	g_{\r}(B',B'')  &= \frac{1}{2} \text{Tr} \left( \r \{ B', B'' \}\right)\\ &= \text{Re} (\xi',\xi'').
\end{align}
This discussion essentially shows that the two geometrical structures in the ray space $\mathcal{R}$, the symplectic structure and the Riemannian structure, arise from the inner product of two horizontal vectors, and can be expressed in terms of two tangent vectors $B',B'' \in T_{\r}\mathcal{R}$ as:
\begin{align} \label{twogeo}
\text{Tr} \left( \r B' B''\right) &= (\xi', \xi'') \\&= g_{\r}(B',B'') + \frac{i}{2} \omega_{\r}(B',B''). 
\end{align}
It turns out that the relation between the two geometrical structures is much deeper and intimate, and will be evident in the next subsection.

\subsection{K\"{a}hler structure of ray space}

Any vector $\j \in \mathcal{B}$ can be completely specified by the N-tuple of complex numbers $\{ x_{1}, x_{2}, \cdots, x_{N} \}$, which are the complex coefficients $x_j = ({e}_j,\j)$ of the expansion of $\j$ in any orthonormal basis $\{ e_{j}\}$ that spans $\mathcal{H}$. Henceforth for simplicity we shall assume that $N$ is finite. The state space $\mathcal{B}$ is thus isomorphic to a subset $\mathcal{S}$ of $\cc^{N}$ wherein the normalisation condition: $\sum_{j=1}^{N} |x_{j}|^{2} = 1$ is respected. 

As defined earlier, two vectors belong to the same ray if they differ by a complex multiple. This fact can now be expressed in terms of N-tuples: two N-tuples differing with an overall complex multiple are identified $(x_{1}, x_{2}, \cdots, x_{N}) \sim ( \l x_{1}, \l x_{2}, \cdots, \l x_{N})$, for any non-zero complex number $\l$. The ray space correspondingly obtained is the well known\emph{Complex Projective Space} $\cc \mathrm{P}^{N-1} \equiv \mathcal{S}/\sim$ \cite{nakahara2003,mukunda1993,arnobohm,page1987,kobayashi1996}.  It can be seen that the ray space $\mathcal{R}$ and $\cc \mathrm{P}^{N-1}$ are isomorphic. The significance of this identification is that it allows one to think of ray space $\mathcal{R}$ as a complex analytic manifold \cite{nakahara2003,eguchi1980gravitation,kobayashi1996}.

In a neighbourhood of $\mathcal{S}$ wherein $x_{N} \neq 0$ one can define set of $N-1$ complex numbers:
\begin{align}
	z^{i} = \frac{x_{i}}{x_{N}},
\end{align}
for $i < N$, which are called inhomogeneous coordinates \cite{nakahara2003,page1987,kobayashi1996}. Evidently these coordinates are immune to the redefinition $x_{i} \rightarrow \l x_{i}$ and hence  provide a coordinate system over $\cc \mathrm{P}^{N-1}$. The tangent space $T_{P}\cc \mathrm{P}^{N-1}$ is each point of $\cc \mathrm{P}^{N-1}$ is spanned by the tangent vectors $\{ \frac{\d}{\d z^{1}} , \frac{\d}{\d z^{2}} , \cdots, \frac{\d}{\d \bar{z}^{1}}, \frac{\d}{\d \bar{z}^{2}}, \cdots \}$. Corresponding to these basis vectors, there exists a set of one-forms $\{ d z^{1}, d z^{2}, \cdots, d \bar{z}^{1}, d \bar{z}^{2}, \cdots\}$, which form the basis of the cotangent space $T_{P}^{\ast} \cc \mathrm{P}^{N-1}$. Owing to the dual nature of these forms, we have the following relations due to their interior contraction:
\begin{align}
&	i_{\frac{\d}{\d \bar{z}^{i}}} d z^{j} = \langle d z^{j}, \frac{\d}{\d \bar{z}^{i}} \rangle = \del_{i}^{j}, \\ & i_{\frac{\d}{\d \bar{z}^{i}}} d \bar{z}^{j} = \langle d \bar{z}^{j}, \frac{\d}{\d {z}^{i}} \rangle = \del_{i}^{j},\\
&	i_{\frac{\d}{\d {z}^{i}}} d \bar{z}^{j} = 0 = i_{\frac{\d}{\d \bar{z}^{i}}} d {z}^{j} = 0.
\end{align} 
The projective space is also endowed with the existence of a linear map $J$ called the \emph{almost complex structure}\cite{kobayashi1996,nakahara2003,eguchi1980gravitation} whose action on the basis is given by:
\begin{align} \label{com1}
	J(\frac{\d}{\d z^{i}}) &= i  \frac{\d}{\d z^{i}},\\ \label{com2}
    J(\frac{\d}{\d \bar{z}^{i}}) &= -i  \frac{\d}{\d \bar{z}^{i}}.
\end{align}

Now suppose one is given a closed curve $C$ in some neighbourhood of $\mathcal{B}$ specified as:
$C = \{ \j(s) \in  \mathcal{B} \lvert \: 0 \leq s \leq \L, \: \text{and}\:\j(0) = \j(\L)\}$.
Then corresponding to such a curve one can also define a closed curve $\mathscr{C}$ in $\mathcal{S}$. The connection functional $A_{\j}(s)$ defined using $\j(s)$, which reads $A_{\j}(s) = \frac{\text{Im} (\j, \j_{s})}{(\j,\j)}$, can also be correspondingly defined over the closed curve $\mathscr{C}$ in $\mathcal{S}$ and it reads:
\begin{align}
A_{\j}(s) = \frac{1}{2 i \sum_{j=1}^{N} \bar{x}^{j} x^{j}}	\sum_{j=1}^{N} \left( \bar{x}^{j} \dot{x}^{j} - \dot{\bar{x}}^{j} x^{j}  \right). 
\end{align}
Here the dot stands for the differentiation with respect to $s$. 
The geometric phase corresponding to curve $C$, which is $\gg[C] = \oint_{C} \: ds \: A_{\j(s)}(s)$ can now be expressed as a closed integral over the curve $\mathscr{C}$:
\begin{align}
\gg[C] &= \oint_{\mathscr{C}} \: ds \: 	\frac{1}{2 i \sum_{j=1}^{N} \bar{x}^{j} x^{j}}	\sum_{j=1}^{N} \left( \bar{x}^{j} \dot{x}^{j} - \dot{\bar{x}}^{j} x^{j}  \right). 
\end{align} 
This closed loop integral can be expressed as an integral over one-form $A$ as $\gg[C] = \oint_{\mathscr{C}} \: A$, where $A$ is given by:
\begin{align}
	A = \frac{1}{2 i} d\left( \ln \frac{x^{N}}{\bar{x}^{N}}\right) + \frac{1}{2 i} 
	\frac{\sum_{j=1}^{N-1} \left( \bar{z}^{j} d z^{j} - z^{j} d \bar{z}^{j} \right)}{\left( \sum_{j=1}^{N-1} |z^{j}|^{2} + 1\right)}.
\end{align}
We immediately see that owing to the presence of the first term, this one-form is not projectable to the ray space. Instead if we consider the two-form $F = dA$ then the contribution due this term vanishes, and one obtains:
\begin{align}
	F = \sum_{j,k=1}^{N-1} F_{j \bar{k}} \left(dz^{j} \wedge d\bar{z}^{k} \right).
\end{align}
Here the complex coefficients $F_{j \bar{k}}( = - F_{\bar{k} j})$ specify the action of $F$ on the tangent vectors 
$F_{j \bar{k}} = F(\frac{\d}{\d {z}^{j}}, \frac{\d}{\d \bar{z}^{k}})$ and are explicitly given by:
\begin{align}
F_{j \bar{k}} = i \frac{\del_{jk}   \left( \sum_{l=1}^{N-1} |z^{l}|^{2} + 1 \right) - \bar{z}^{j} z^{k} }{\left( \sum_{l=1}^{N-1} |z^{l}|^{2} + 1 \right)^2}.	
\end{align}
This defines a symplectic structure on the projective space $\cc \mathrm{P}^{N-1}$. The geometric phase can now be alternatively determined by the surface integral $\gg[C] = \int_{\mathscr{S}} dA$ of this two-form $F = dA$ which is defined over the surface $\mathscr{S}$ of $\cc \mathrm{P}^{N-1}$ which is bounded by the image of $\mathscr{C}$.

There also exists a Riemannian metric structure $g$ on the projective space, which is compatible with the symplectic structure $F$ and the almost complex structure $J$. To see this consider a positive semi-definite symmetric bilinear map $g$ that maps a pair of tangent vectors $X$ and $Y$ to non-negative reals which is defined using the two-form $F$ and linear map $J$ as:
\begin{align}
	g(X,Y) = F(X, J Y). 
\end{align} 
From here it follows that:
\begin{align}
g = \sum_{j,k=1}^{N-1} 2 g_{j \bar{k}} dz^{j} d\bar{z}^{k}, 
\end{align}
where the coefficients are given  by:
\begin{align}
g_{j \bar{k}} = - i F_{j \bar{k}} = \frac{\del_{jk} \left( \sum_{l=1}^{N-1} |z^{l}|^{2} + 1 \right) - \bar{z}^{j} z^{k} }{\left( \sum_{l=1}^{N-1} |z^{l}|^{2} + 1 \right)^2}.
\end{align}
This metric is the well known \emph{Fubini-Study metric} on the complex projective space\cite{nakahara2003,page1987,anandan1988geometric,eguchi1980gravitation,samuel1,kobayashi1996}.

Interestingly the coefficients $g_{j \bar{k}}$ can be obtained from the differentiation of an object $\mathscr{K}$ known as the \emph{K\"ahler potential}:
\begin{align}
g_{j \bar{k}} = \frac{\d^2 \mathscr{K}}{\d z^{j} \d \bar{z}^{k}},
\end{align}
where 
\begin{align}
	\mathscr{K} = \ln \left( \sum_{l=1}^{N-1} |z^{l}|^{2} + 1 \right).
\end{align}
In the literature, a manifold wherein the Riemannian metric $g$ co-exists with a closed two-form $F$
via the almost complex structure $J$ is known as a \emph{K\"ahler manifold} \cite{nakahara2003,page1987,eguchi1980gravitation}. From the above treatment, we see that not only the complex projective space $\cc \mathrm{P}^{N-1}$ is a K\"ahler manifold, but the ray space $\mathcal{R}$ can also be thought of as being a K\"ahler manifold. The origin and interconnection of the Riemannian and symplectic structures on the ray space is now clearly visible, as also its relevance for the existence of geometric phase.

\section{Geometry of Generalised Geometric Phase \label{ggp}}

Having briefly reviewed the mathematical structures that are responsible for the origin of the usual geometric phase, we shall now study the framework that is needed to understand the reason for the existence of the generalised geometric phase.

\subsection{Fibre bundle structures}

Let us assume that we are given a quantum system at hand which is defined over a Hilbert space $\hh$ whose complex dimension is $N$, which may nor may not be finite. Over this Hilbert space let us consider an observable $\oo$ which is a self-adjoint linear operator. We shall also assume that $\oo$ admits distinct real eigenvalues $\l_{i}$ such that: 
\begin{align} \nmb
\oo e_{i} = \l_{i} e_{i},
\end{align}
where eigenvectors $e_{i} \in \hh$ (where $i=1,2,\cdots,N$) are orthonormal, 
\begin{align} \nmb
(e_{i},e_{j})= \del_{ij}.
\end{align}
For any given non-zero vector $\j \in \hh$, the \emph{quantum average or the expectation value} $O_{\j}$ of the observable $\oo$ corresponding to the state $\j$ is a real number given by:
\begin{align} \nmb
O_{\j} = \frac{(\j, \oo \j)}{(\j,\j)}.
\end{align}
It is well known that the quantum average lies between the smallest eigenvalue $\l_{min}$ and the largest eigenvalue $\l_{max}$ of $\oo$: $\l_{min} \leq O_{\j} \leq \l_{max}$.  

In the usual treatments, the state $\j$ under consideration is generally assumed to be unit normalised. In this discussion instead we fix the norm of $\j$ as:
\begin{align} \label{c1}
(\j, \oo \j) & = 1, \text{ \: if } O_{\j} > 0; \\
 \label{c2}
(\j, \oo \j) &= -1, \text{ if } O_{\j} < 0.
\end{align} 
This normalisation condition can also be expressed as $(\j,\j) = |O_{\j}|^{-1}$. Evidently this condition is ill-defined when $(\j, \oo \j) = 0$. Let us to define the set $\mathcal{K}_{\oo}$ which consists of all such states:
\begin{align}
	\mathcal{K}_{\oo} = \{ \f \in \hh \: \lvert \: (\f, \oo \f ) = 0.   \}.
\end{align}
It may be noted that $\mathcal{K}_{\oo}$ does not form a subspace of $\hh$. However it does have an interesting mathematical structure, it is a homogeneous quadratic form. Resolving the state $\f$ into eigenstates $e_{i}$, allows one to express the equation $(\f, \oo \f ) = 0$ as:
\begin{align} \label{ksurface}
	\sum_{j=1}^{N} \l_{j} |w_{j}|^2 = 0,
\end{align} 
where $w_{j} = (e_{j},\f)$. This relation shows that $\mathcal{K}_{\oo}$ is a homogeneous quadric surface, which is a generalisation of conical surface to higher dimensions, embedded into $\rr^{2N}$ and of real dimension $2r - 1$, where $r$ is the rank of $\oo$.

In this discussion, we shall only be considering the vectors that belong to $\hh$ but are not in $\mathcal{K}_{\oo}$, that is, which belong to the set $\ho = \hh - \mathcal{K}_{\oo}$. Any state in $\ho$ can always be normalised so that $(\j, \oo \j) = \text{sgn}\: O_{\j}$, and can be brought to either of the sets $\no^{\pm}$, where: 
\begin{align}
\no^{\pm} = \{ \f \in \ho \lvert \; \text{where} \; (\f, \oo \f) = \pm 1 \}.
\end{align}
The sets $\no^{\pm}$ so defined are disjoint sets and none of them form a linear vector space.  In terms of the complex coefficients $w_{j} = (e_{j},\j)$ the above two conditions (\ref{c1}) and (\ref{c2}) read:
\begin{align} \nmb
\sum_{j=1}^{N} \l_{j}|w_{j}|^{2} = \pm 1.
\end{align}
This relation shows that the sets $\no^{\pm}$ are smooth quadric surfaces of real dimension $2r - 1$ embedded in $\rr^{2N}$. It is well known that the quadric surfaces are a generalisation of the conic sections into higher dimensions. In case when all the eigenvalues $\l_{j}$ are either positive or negative, only one of the manifolds from $\no^{\pm}$ exists, and is \emph{a generalised ellipsoid}. In the general case, when the eigenvalues need not obey such a restriction, the manifolds $\no^{\pm}$ both can exist and are found to be \emph{generalised hyperboloids}. 

One can now define the pair of ray spaces $\ro^{\pm}$ as projective spaces $\ro^{\pm} = \no^{\pm}/\sim$; wherein any two states, respectively in $\no^{\pm}$, differing by a phase are identified: $\j \sim \: e^{i \q} \j \in \no^{\pm}$. This essentially defines a projection map $\pi : \no^{\pm} \ra \ro^{\pm}$ that maps the vectors to density matrices: 
\begin{align}
	\ro^{\pm} = \left\lbrace  \r = \j \j^{\dagger} \bigg| \: \j \in \no^{\pm} \right\rbrace.
\end{align}
It must be noted that these density matrices are defined such that: $\r = \r^{\dagger}$, $\r^2 = \r \: \text{Tr} \r$, and $\text{Tr} \r > 0$.

Thus one sees that the pair of triples $(\no^{+},\ro^{+},\pi)$ and $(\no^{-},\ro^{-},\pi)$ form separate principle fibre bundles respectively with the total space as $\no^{\pm}$, base space as $\ro^{\pm}$, with $U(1)$ as the structure group\cite{mukunda1993,samuel1,bertlmann2000anomalies,anandan1988geometric,bsimon1983}.
It must be understood that these ray spaces $\ro^{\pm}$ are completely different from the ray space $\mathcal{R}$, that is formed by projecting the vectors that belong to the set $\mathcal{B}$.

\begin{figure}
	\begin{center}
		\includegraphics[scale=0.35]{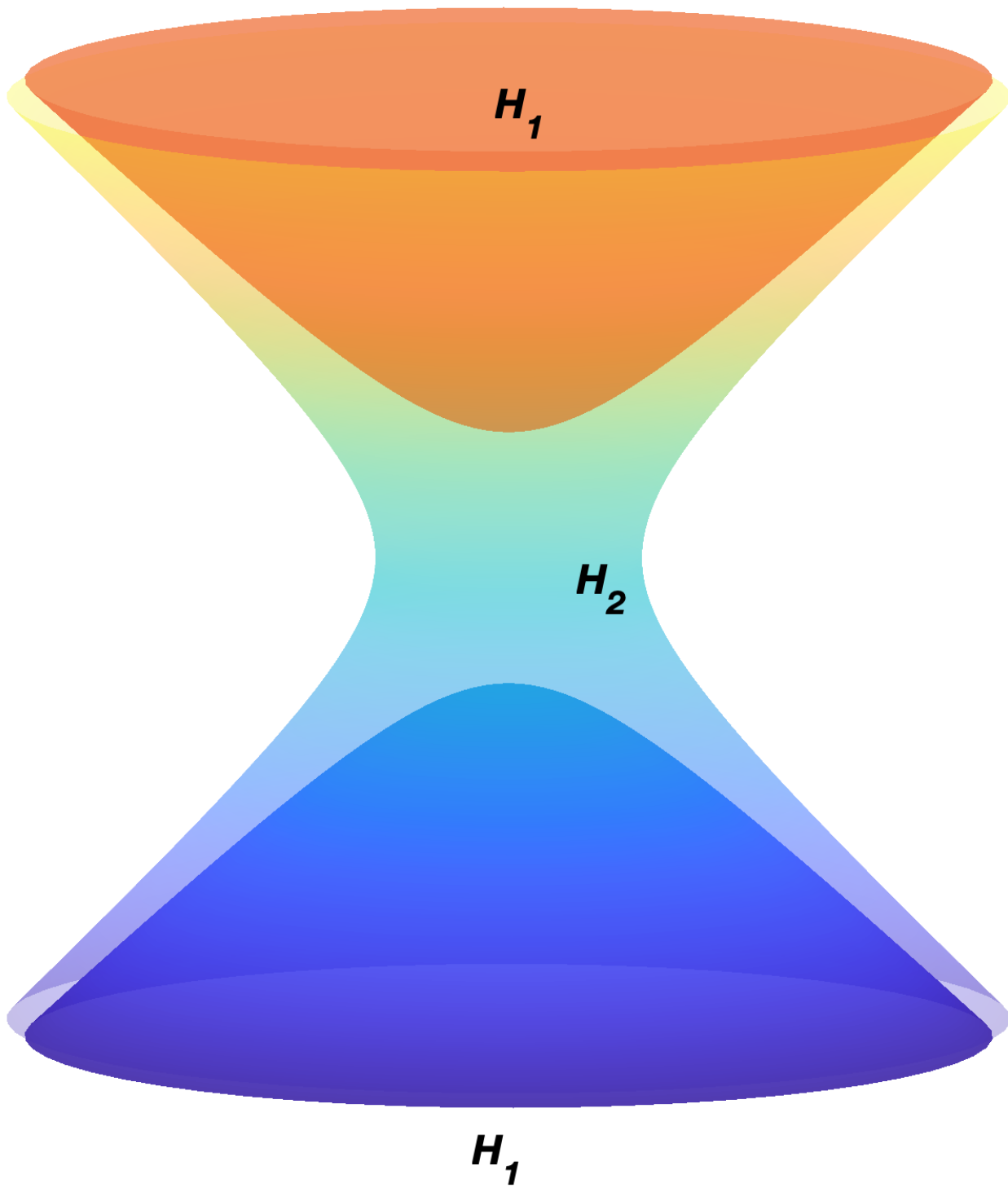}\caption{A schematic representation of the ray spaces for two state system when $\l_j = \pm 1$. Here the two-sheeted hyperboloid $H_{1}$ represents the ray space manifold $\ro^+$, whereas the one-sheeted hyperboloid $H_{2}$ is depicting the ray space manifold $\ro^-$.  \label{fig1}}	
	\end{center}
\end{figure}

In order to better understand these ideas, let us consider an example of a two state quantum system, for which dimension of $\hh$ is 2, and an observable $\oo$ such that it admits two eigenvalues $\pm 1$. The total spaces $\no^{\pm}$ are defined by the relation:
\begin{align}
	|w_1|^2 - |w_2|^2 = \pm 1.
\end{align}
Writing $w_{1,2} = r_{1,2} e^{i \q_{1,2}}$, we see that this equation is obeyed for all the choices of $\q_{1,2}$ so long as $r_1^2 - r_2^2 = \pm 1$. Let us set $w_1 = r_1$ and $w_2 = r_2 e^{i\f}$, with $\f  = \q_2 - \q_1$ being the relative phase. By these choices, we have gotten rid of the global $U(1)$ phase freedom, and so $w_{1,2}$ can now be thought of as the coordinates of the ray spaces $\ro^{\pm}$. Writing $w_2 = u_2 + i v_2$, the above relation now reads:
\begin{align}
	r^{2}_{1} - u^{2}_{2} - v^{2}_{2} = \pm 1. 
\end{align} 
This shows that the ray space manifold $\ro^{+}$ is a \emph{two-sheeted hyperboloid}. It can be explicitly parametrised in terms of two angles $\q$ and $\f$ as: $r_1 = \pm \cosh (\q/2)$, $u_2 = \sinh (\q/2) \cos \f$, $v_2 = \sinh (\q/2) \sin \f$. Whereas the ray space manifold $\ro^{-}$ is a \emph{one-sheeted hyperboloid}, parametrised  as $r_1 = \sinh (\q/2)$, $u_2 = \cosh(\q/2) \cos \f$, $v_2 = \cosh (\q/2) \sin \f$. Both these manifolds are depicted graphically in Fig. \ref{fig1}. This treatment clearly shows that the two ray spaces $\ro^{\pm}$ can be of completely character from one another, although they are capturing the physical aspects of the same underlying quantum system. The states for which $w_1 = w_2$ belong to $\mathcal{K}_{\oo}$. It is worth mentioning that in this physical system, the ray space $\mathcal{R}$ is a two sphere $S^2$ popularly known as the Bloch sphere \cite{mukunda1993,agarwal2013,nakahara2003}, which is fundamentally different from the ray spaces $\ro^{\pm}$.     

Having found the existence of the pair of ray spaces $\ro^{\pm}$, one can anticipate the occurrence of geometrical structures on them, as in the case of $\mathcal{R}$. In order to obtain further insight in this direction, let us define the tangent space $T_{\j} \no^{\pm}$ corresponding to vector $\j \in \no^{\pm}$ as:
\begin{align}
T_{\j} \no^{\pm} = \left\lbrace \f \in \ho \bigg| \text{Re}(\j , \oo \f )  = 0 \right\rbrace.	
\end{align}
It follows from here that the tangent spaces $T_{\j} \no^{\pm}$ form real linear vector space of dimension $2N-1$. The vertical subspace $V_{\j} \no^{\pm} \subset T_{\j} \no^{\pm}$ is given by:
\begin{align}
V_{\j} \no^{\pm} = \left\lbrace i \a \j \bigg| \a \in \rr \right\rbrace. 
\end{align}
Keeping in mind the discussions in the Introduction on the generalised geometric phase, we define the \emph{generalised connection} on the manifolds $\no^{\pm}$ at some point $\j$ as a linear functional on the tangent space $T_{\j} \no^{\pm}$ which is given by:
\begin{align}
A^{\oo}_{\j}(\f) &=  \text{Im}(\j , \oo \f ), \text{for any}\: \f \in T_{\j} \no^{+},\\
 &=  -\text{Im}(\j , \oo \f ), \text{for any}\: \f \in T_{\j} \no^{-}.
\end{align}
Using this we have the horizontal subspaces $H_{\j} \no^{\pm}$ as:
\begin{align}
	H_{\j} \no^{\pm} = \left\lbrace \f \in T_{\j} \no^{\pm} \bigg|  A^{\oo}_{\j}(\f) = 0 \right\rbrace. 
\end{align}
This subspace is a real linear vector space of dimension $2(N-1)$. It can be readily checked that any tangent vector can be uniquely into components belonging to vertical and horizontal subspaces, so that $T_{\j} \no^{\pm} = H_{\j} \no^{\pm} \oplus V_{\j} \no^{\pm}$.  
 
In the forthcoming discussion, in the favour of simplicity, we shall consider the case when one is working with states defined in $\no^{+}$. However it will be eventually clear that the similar construction will also hold when one is dealing with the vectors defined in $\no^{-}$. 

Let us consider some fixed point $\r_{0} \in \ro^{+}$, defined through a fixed vector $\j_{0} \in \no^{+}$. We choose to work in a connected neighbourhood in $\ro^{+}$ around $\r_{0}$, through the vector $\j(\c_{0},\a)$:
\begin{align}
\j(\c_{0},\a)  = e^{i \a} \left(  c \j_{0} + \c_{0} \right),
\end{align}  
where $c = \sqrt{1 - (\c_{0}, \oo \c_{0})}$, for $\c_{0} \in H_{\j_0} \no^{+}$ with the condition $(\c_{0}, \oo \c_{0}) < 1$. Evidently the neighbourhood is coordinatised by the vector $\c_{0}$ and $\a$. Under infinitesimal changes in the coordinates, the vector $\j(\c_{0},\a)$ changes as:
\begin{widetext}
\begin{align}
	\del \j(\c_{0},\a) = i \del \a \j(\c_{0},\a) + e^{i \a} \del \c_{0} - \frac{e^{i \a}}{c} \text{Re} (\del \c_{0}, \oo \c_{0}) \j(\c_{0},\a). 
\end{align}
\end{widetext}
If this changed vector lies in the tangent space then one has $\del \j(\c_{0},\a) = \varepsilon (i a \j(\c_{0},\a) + \c)$, for some $a \in \rr$ and $\c \in H_{\j} \no^{+}$. Here $\varepsilon$ is a real infinitesimal. From here it follows that:
\begin{align}
&    \del \c_{0} = \varepsilon \left[ \c + \frac{\j_{0}}{c} \text{Re} (\c_0, \oo \c) + \frac{i}{c^2} \text{Im} (\c_0, \oo \c) \j \right], \\
& \del \a = \varepsilon \left[ a - \frac{1}{c^2} \text{Im} (\c_0, \oo \c)\right].
\end{align}
These two relations allows us to have a differential geometric representation of the tangent vector $\f = (i a \j + \c) \in T_{\j} \no^{+}$, with $\a$ and $\c_0$ as the coordinates, as:
\begin{align} \label{tvecrep}
X_{\f} = \frac{\del \a}{\varepsilon} \frac{\d}{\d \a} + \left( \frac{\del \c_0}{\varepsilon}\right)^{\dagger} \frac{\d}{\d \c_0^{\dagger}} + \frac{\d}{\d \c_0} \left( \frac{\del \c_0}{\varepsilon} \right). 
\end{align}
We can now have a differential geometric representation of the generalised connection as a one-form $A^{\oo}$ which could be expressible as:
\begin{align} \nmb
    A^\oo = \kappa \frac{\d}{\d \a} + \e^{\dagger} d \c_0 + d \c_{0}^{\dagger} \e,
\end{align}
for some undetermined $\kappa$ and $\e$. These two unknowns can be determined by demanding that the interior product of the connection one-form with the tangent vector $X_\f$ must agree with its functional representation:
\begin{align}
    i_{X_{\f}} A^{\oo} = A^{\oo}_{\j}(\f) = \text{Im} (\j, \oo \f).
\end{align}
This condition fixes the unknowns as $\kappa = 1$ and $\e = \frac{i}{2} \oo \c_0$, so that the connection one-form defined at point $\j(\c_0,\a)$ is explicitly given by:
\begin{align} \label{cononef}
    A^{\oo} = d \a - \frac{i}{2} \left( ( \oo \c_0)^{\dagger} d \c_0 - d \c^{\dagger}_0 (\oo \c_0) \right).
\end{align}
This expression takes an elegant form by employing the coefficients $w_{j} = (e_{j},\c_{0})$ of the vector $\c_0$ in the eigenbasis of $\oo$, and its reads:
\begin{align}
    A^{\oo} = d \a - \frac{i}{2} \sum_{j=1}^{N} \l_{j} \left( w^{\ast}_{j} d w_{j} - d w^{\ast}_{j} w_{j} \right).
\end{align}
Thus one sees that the one-form $A^{\oo}$ is a generalisation of one-form $A$ given by (\ref{forma}). The presence of the term $d \a$ in the above expression clearly indicates that the one-form $A^{\oo}$ can not be expressed as a pullback via $\pi^\ast$ for any one-form defined over $\ro^{+}$. 

Consider a closed curve $\mathcal{C}$ specified by states $\j(\c_{0}(l),\a(l))$ in a neighbourhood of $\no^{+}$ which is parameterised by a real parameter $0\leq l \leq \L$,  and is  such that $\j(\c_{0}(0),\a(0)) \equiv \j(\c_{0}(\L),\a(\L))$. The generalised geometric phase for such a curve is the (an)holonomy of the connection one-form along the curve \cite{vyas2023}: 
\begin{align} \nmb
\gg_{\oo}[\mathcal{C}] &= \oint_{\mathcal{C}} \: dl \: A^{\oo}_{\j}(\dot{\j}(l)) 
= \oint_{\mathcal{C}} \: dl \: \text{Im} (\j(l),\oo \dot{\j}(l)) \\
&= \oint_{\mathcal{C}} \: A^{\oo}. 	
\end{align}
One can now convert this loop integral into a surface integral of the two-form $d A^{\oo}$ over a smooth surface $\mathcal{S}$ in $\no^{+}$ whose boundary is $\mathcal{C}$:
\begin{align}
\gg_{\oo}[\mathcal{C}] = \int_{\mathcal{S}} \: d A^{\oo}. 	
\end{align}
The two form $d A^{\oo}$ defined over $\no^{+}$ is independent of $\a$, and is given by:
\begin{align}
    dA^{\oo} &= - i \: d\c^{\dagger}_{0} \wedge \oo d \c_{0}.
\end{align}
Being independent of $\a$ allows this two-form to be expressed as pullback via $\pi^{\ast}$ of a two-form $\omega^{\oo}$ defined over $\ro^+$:
\begin{align}
	dA^{\oo} = \pi^{\ast} \omega_{\r}^{\oo}, 
\end{align}
so that the geometric phase can be written as $\gg_{\oo}[\mathcal{C}] = \int_{\pi{(\mathcal{S})}} \: \omega^{\oo}_{\r}$. Since the rays in $\ro^+$ depend only on $\c_{0}$ and $\c^{\dagger}_{0}$, one can think of these two horizontal vectors as coordinates on $\ro^+$. This allows us to define the two-form as: $\omega^{\oo}_{\r} = - i \: d\c^{\dagger}_{0} \wedge \oo d \c_{0}$.  
Evidently this two-form is closed but not exact, and it defines a symplectic structure on $\ro^+$. It can be expressed locally, using the coordinates $w_{j} = \frac{1}{\sqrt{2}} (q_{j} - i p_{j})$ (where $w_{j} = (e_{j}, \c_{0})$) as:
\begin{align}
	\omega_{\r}^{\oo} = \sum_{j=1}^{N} \l_{j} \: dp_{j} \wedge dq_{j}. 
\end{align}
This representation makes it clear that this symplectic two-form is a generalisation of the $\omega_{\r}$ given by (\ref{twoform}), and $\omega_{\r}^{\oo}$ goes over to $\omega_{\r}$ when $\oo \equiv \mathrm{1}$.

Let us have two tangent vectors $\f_{1}, \f_{2} \in T_{\j} \no^{+}$, and their corresponding differential geometric representation $X_{\f_{1,2}}$ as in (\ref{tvecrep}). Performing interior product of $dA^{\oo}$ with these two tangent vectors yields:
\begin{align} \nmb
    dA^{\oo}_{\j}(\f_1, \f_2) &= i_{X_{\f_1}} i_{X_{\f_1}} dA^{\oo} = 2 \: \text{Im} \: (\f_1, \oo \f_2) \\ &= 2 \: \text{Im} \: (\c_1, \oo \c_2)
\end{align}
which is the corresponding bilinear antisymmetric functional representation of the two-form $d A^{\oo}$. Here $\c_1$ and $\c_2$ respectively represent the horizontal components of the tangent vectors $\f_{1}$ and $\f_{2}$. 

This representation can also be simply obtained by explicitly considering the surface integral $\int_{\mathcal{S}} \: d A^{\oo}$ over a two dimensional surface $\mathcal{S} \in \no^{+}$. Let this surface be given by the set of states $\j(u,v)$, which are parameterised by two coordinates $(u,v)$ and be such that it is bounded by $\mathcal{C}$. The generalised geometric phase in such a case reads:
\begin{widetext}
\begin{align} \nmb
	\gg_{\oo}[\mathcal{C}] & = \oint_{\mathcal{C}} \: dl \: \text{Im} (\j(l), \oo \dot{\j}(l))\\
\nmb	& = \text{Im} \int_{\mathcal{S}} \: du dv \: \left[ \frac{\d}{\d u} \left( \j(u,v), \oo \j_{v}(u,v)\right) - \frac{\d}{\d v} \left( \j(u,v), \oo \j_{u}(u,v)\right) \right] \\
	& = \int_{\mathcal{S}} \: du dv\: 2 \; \text{Im} \: (\j_u, \oo \j_v).
\end{align}
\end{widetext}	
Here the tangent vectors $\j_{u}$ and $\j_{v}$ are derivatives of $\j(u,v)$ respectively with $u$ and $v$.

It is rewarding to find a representation of $\omega_{\r}^{\oo}$ solely in terms of the ray space objects. To do so let us consider a smooth curve $\j(l) \in \no^{+}$, to which there exists a corresponding curve $\r(l) = \j(l) \j^{\dagger}(l) \in \ro^{+}$. At any point $l=q$ the tangent vector $B_{\j}(q)$ is given by:
\begin{align}
B_{\j}(q) = \dot{\r}_{\j}(q) = \dot{\j}(q) \j^{\dagger}(q) + \j(q) \dot{\j}^{\dagger}(q).
\end{align}
We thus see that the tangent vectors at any point $\r$ belong to the tangent space $T_{\r} \ro^{+}$ which is defined as:
\begin{align}
	T_{\r} \ro^{+} = \{ B = \c \j^{\dagger} + \j \c^{\dagger} | \c \in H_{\j} \no^{+}  \}.
\end{align}
In terms of any two tangent vectors $B'$ and $B''$ that belong to $T_{\r} \ro^{+}$, which are obtained from horizontal vectors $\c'$ and $\c''$ respectively, the two-form $\omega_{\r}^{\oo}$ can be expressed as a bilinear antisymmetric functional:
\begin{align} \label{omegaor}
	\omega_{\r}^{\oo}(B',B'') &= - i \: \text{Tr} \left[ \r \oo \left( B' \oo B'' - B'' \oo B' \right) \oo \right] \\
	&=2 \: \text{Im} (\c',\oo \c'').
\end{align}  
One has thus succeeded in expressing the two-form $\omega_{\r}^{\oo}$ in a language that is intrinsic to the ray space $\ro^{+}$ without resorting to the corresponding state space $\no^{+}$.

This treatment emphatically shows that the ray space $\ro^{+}$ is endowed with a symplectic structure which is fundamentally different from the one found on $\mathcal{R}$. And it is this symplectic structure that is responsible for the existence of the generalised geometric phase.

Having found the symplectic structure, one wonders if the ray space $\ro^{+}$ also possesses a Riemannian metric structure, as found over $\mathcal{R}$. To throw light on this aspect, consider a smooth curve in a connected neighbourhood of $\no^{+}$ specified by vector $\j(s)$ and parametrised by $s$. We are interested in providing a notion of distance between the two vectors $\j(s)$ and $\j(s + ds)$  due to infinitesimal change $ds$. Conventionally one works with the squared norm of the difference $(\j(s + ds) - \j(s))$ to give a sense of distance $dl^2$ between the two states. However one can generalise this notion and define the distance via the quantum average of the difference $(\j(s + ds) - \j(s))$ with $\oo$. This can be expressed as:
\begin{align} \nmb
	dl^2 &= \left(\j(s + ds) - \j(s),\oo(\j(s + ds) - \j(s))\right) \\ \nmb
	&= \left(\frac{d \j}{ds}, \oo \frac{d \j}{ds} \right) ds^2. 
\end{align}
Evidently this distance is not projectable to $\ro^{+}$, and can not be interpreted as a distance between $\pi(\j(s))$ and $\pi(\j(s + ds))$. To do so one needs to employ the covariant derivative of $\j$ defined using $A^{\oo}$:
\begin{align}
	\frac{D \j(s)}{ds} = \frac{d \j(s)}{ds} - i A^{\oo}_{\j}(s) \j(s),
\end{align}
where $A^{\oo}_{\j}(s) = \text{Im} (\j, \oo \frac{d \j}{ds})$. So one can now define a notion of distance $d\mathscr{L}^2$ between $\pi(\j(s))$ and $\pi(\j(s + ds))$ as:
\begin{align} \label{distt}
	d\mathscr{L}^{2}(\pi(\j(s)),\pi(\j(s + ds))) = \left(\frac{D \j}{ds}, \oo \frac{D \j}{ds} \right) ds^2.
\end{align}
This shows the existence of a \emph{pseudo-Riemannian metric} $g_{\oo}(\c',\c'')$ on the ray space $\ro^{+}$  which is an \emph{indefinite} symmetric bilinear functional of two horizontal vectors $\c',\c'' \in H_{\j} \no^{+}$, corresponding to point $\j \in \no^{+}$, defined as:
\begin{align} \label{metricor}
	g_{\oo}(\c',\c'') &= \text{Re} \left( \c',\oo \c''\right) \\
	&=\frac{1}{2} \text{Tr} \left[ \r \oo \left( B' \oo B'' + B'' \oo B' \right) \oo \right].
\end{align} 
Notice that this metric on the ray space $\ro^{+}$ is fundamentally different than the metric $g$ defined on $\mathcal{R}$ that is given in (\ref{metric}).

From (\ref{omegaor}) and (\ref{metricor}) one sees that the symplectic structure and the pseudo-Riemannian structure in the ray space $\ro^{+}$ essentially arises from the transition matrix element $\left( \c',\oo \c''\right)$ of two horizontal vectors through the observable $\oo$:
\begin{align}
	\text{Tr} \left[ \r \oo B' \oo B'' \oo \right] &= \left( \c',\oo \c''\right) \\ &= g_{\oo}(B',B'') + \frac{i}{2}\omega^{\oo}_{\r} (B',B'').
\end{align} 
This is a very important relation, since it shows how the new generalised geometrical structures have arisen owing to the change in the notion of the relative phase, which involves the matrix element of two vectors with $\oo$, as discussed in Ref. \onlinecite{vyas2023}.

It is worth noting that the above defined distance can be written for any two rays $\pi(\j_1)$ and $\pi(\j_2)$ as:
\begin{align}
	D\mathscr{L}^{2}(\pi(\j_2),\pi(\j_1)) &= 1 - (\j_1, \oo \j_2 ) (\j_2, \oo \j_1 ) \\
	& = 1- \text{Tr} (\r_{1} \oo \r_{2} \oo),	
\end{align}
where $\r_{1} = \j_1 \j^{\dagger}_1$ and $\r_{2} = \j_2 \j^{\dagger}_2$.

\subsection{Pseudo-K\"ahler structure of the ray space}

Any state $\j$ in the space $\no^{+}$ is completely specified by $N$ complex coefficients \newline $\{w_1, w_2, \cdots, w_N \}$ which are defined as: $w_{j} = (e_{j},\j)$. Further these complex coefficients also respect the relation $\sum_{j=1}^{N} \l_{j} |w_{j}|^{2} = 1$. In this discussion we shall assume that all eigenvalues $\l_{j}$ are non-zero and distinct for simplicity.  The state space $\no^{+}$ is thus isomorphic to a subset $\mathcal{N}^{+}$ of $\cc^{N}$, which is defined as:
\begin{align}
	\mathcal{N}^{\pm} = \left\lbrace (w_1,w_2,\cdots,w_N) \in \cc^{N} \: \biggl\lvert \: \sum_{j=1}^{N} \l_{j} |w_{j}|^{2} = \pm 1 \right\rbrace. 
\end{align}
On the physical grounds we identify two N-tuples which differ by an overall constant, so that $(w_1,w_2,\cdots,w_N) \sim (c w_1, c w_2,\cdots,c w_N)$, where $c$ is some non-zero complex number. One thus obtains a \emph{Projective Space} $\mathscr{P}^{+} = \mathcal{N}^{+}/\sim$, which is isomorphic to the ray space $\ro^{+}$.

Consider a neighbourhood in $\mathcal{N}^+$ wherein $w_N \neq 0$. In such a neighbourhood one can define the $N-1$ complex coordinates as:
\begin{align}
	z^{j} = \frac{w_{j}}{w_N}, 
\end{align}
when $j < N$, which provide a coordinate system on $\mathscr{P}^+$. The tangent space $T_{P}\mathscr{P}^+$ at each point of $\mathscr{P}^+$ is spanned by the tangent vectors $\{ \frac{\d}{\d z^{1}} , \frac{\d}{\d z^{2}} , \cdots, \frac{\d}{\d \bar{z}^{1}}, \frac{\d}{\d \bar{z}^{2}}, \cdots \}$. Corresponding to these basis vectors, one defines a set of one-forms $\{ d z^{1}, d z^{2}, \cdots, d \bar{z}^{1}, d \bar{z}^{2}, \cdots\}$, which form the basis of the cotangent space $T_{P}^{\ast} \mathscr{P}^+$. It is evident that the projective space $\mathscr{P}^+$
is also endowed with an almost complex structure $J$, as defined in (\ref{com1}) and (\ref{com2}).

Consider that one is given a closed curve $C$ in some connected neighbourhood of $\no^{+}$ specified as:
$C = \{ \j(s) \in  \no^{+} \lvert \: 0 \leq s \leq \L, \: \text{and}\:\j(0) = \j(\L)\}$.
Then corresponding to such a curve there also exist a closed curve $\mathscr{C}$ in $\mathcal{N}^+$. The connection functional $A^{\oo}_{\j}(s)$ which is defined using $\j(s)$: $A^{\oo}_{\j}(s) = \text{Im} (\j, \oo \j_{s})$, can also be defined over the closed curve in $\mathcal{N}^+$ and it reads:
\begin{align}
	A^{\oo}_{\j}(s) = \frac{1}{2 i \sum_{j=1}^{N} \l_{j} \bar{w}^{j} w^{j}}	\sum_{j=1}^{N} \left( \l_{j} \bar{w}^{j} \dot{w}^{j} - \l_{j} \dot{\bar{w}}^{j} w^{j}  \right). 
\end{align}
Here the dot stands for the differentiation with respect to $s$. 
The geometric phase $\gg_{\oo}[\mathcal{C}] = \oint_{C} \: ds \: A^{\oo}_{\j(s)}(s)$ can now be expressed as a closed integral over the curve $\mathscr{C}$:
\begin{align}
	\gg_{\oo}[C] &= \oint_{\mathscr{C}} \: ds \: 	\frac{1}{2 i \sum_{j=1}^{N} \l_{j} \bar{w}^{j} w^{j}}	\sum_{j=1}^{N} \left( \l_{j} \bar{w}^{j} \dot{w}^{j} - \l_{j} \dot{\bar{w}}^{j} w^{j}  \right). 
\end{align} 
Using the inhomogeneous coordinates $z^{j}$, this loop integral can be expressed as an integral over one-form $A^{\oo}$ as $\gg_{\oo}[C] = \oint_{\mathscr{C}} \: A^{\oo}$, where $A^{\oo}$ is given by:
\begin{align}
	A^{\oo} = \text{Im} \left( \frac{d w^N}{w^N}\right) + \text{Im} \left(
	\frac{\sum_{j=1}^{N-1}  \l_{j} \bar{z}^{j} d z^{j}}{\left( \sum_{j=1}^{N-1}  \l_{j} |z^{j}|^{2} + \l_{N} \right)} \right).
\end{align}
Owing to the presence of the first term, one clearly sees that this one-form is not projectable to the projective space. On the other hand if we construct the corresponding two-form $F^{\oo} = dA^{\oo}$ then the contribution due this term vanishes, and one obtains:
\begin{align}\label{twoformexp}
	F^{\oo} = \sum_{j,k=1}^{N-1} F^{\oo}_{j \bar{k}} \left(dz^{j} \wedge d\bar{z}^{k} \right),
\end{align}
wherein the complex coefficients $F^{\oo}_{j \bar{k}}$ are given by:
\begin{align} \label{twoformoo}
	F^{\oo}_{j \bar{k}} = i \frac{\del_{jk} \l_{j}  \left( \sum_{l=1}^{N-1} \l_{l} |z^{l}|^{2} + \l_{N} \right) - \l_{j} \l_{k} \bar{z}^{j} z^{k} }{\left( \sum_{l=1}^{N-1} \l_{l}|z^{l}|^{2} + \l_{N} \right)^2}.	
\end{align}
The geometric phase can now be alternatively determined by the surface integral $\gg_{\oo}[C] = \int_{\mathscr{S}} dA^{\oo}$ of this two-form $F^{\oo} = dA^{\oo}$ which is defined over the surface $\mathscr{S}$ of projective space which is bounded by the image of $\mathscr{C}$. This shows the existence of a symplectic structure on the projective space $\mathscr{P}^+$. 

Interestingly there also exists a pseudo-Riemannian metric structure $g^{\oo}$ on the projective space, which is compatible with the symplectic structure $F^{\oo}$ and the almost complex structure $J$. Consider an indefinite symmetric bilinear map $g^{\oo}$ that maps a pair of tangent vectors $X$ and $Y$ to reals which is defined using the two-form $F^{\oo}$ and linear map $J$ as:
\begin{align}
	g^{\oo}(X,Y) = F^{\oo}(X, J Y). 
\end{align} 
From \ref{twoformexp} $g^{\oo}$ can be written as:
\begin{align}
	g^{\oo} = \sum_{j,k=1}^{N-1} 2 g^{\oo}_{j \bar{k}} dz^{j} d\bar{z}^{k}, 
\end{align}
where the coefficients are:
\begin{align} \label{metricoo}
	g^{\oo}_{j \bar{k}} = - i F^{\oo}_{j \bar{k}} = \frac{\del_{jk} \l_{j}  \left( \sum_{l=1}^{N-1} \l_{l} |z^{l}|^{2} + \l_{N} \right) - \l_{j} \l_{k} \bar{z}^{j} z^{k} }{\left( \sum_{l=1}^{N-1} \l_{l}|z^{l}|^{2} + \l_{N} \right)^2}.
\end{align}
This metric is a generalisation of the  {Fubini-Study metric} $g$ that was encountered earlier, on the projective space $\cc \mathrm{P}^{N-1}$ \cite{nakahara2003,page1987}. The coefficients of $g^\oo$ and $F^\oo$ can be obtained from the differentiation of the corresponding {K\"ahler potential} $\mathscr{K^{\oo}}$:
\begin{align}
	g^{\oo}_{j \bar{k}} = \frac{\d^2 \mathscr{K}^{\oo}}{\d z^{j} \d \bar{z}^{k}},
\end{align}
where 
\begin{align} \label{kahlerpot}
	\mathscr{K}^{\oo} =  \ln \left( \sum_{l=1}^{N-1} \l_{l} |z^{l}|^{2} + \l_{N} \right).
\end{align}
In this manner we see that the projective space $\mathscr{P}^+$ is the one wherein a pseudo-Riemannian metric $g^{\oo}$ co-exists with a closed K\"ahler two-form $F^{\oo}$
via the almost complex structure $J$, to form a \emph{ pseudo-K\"ahler manifold} \cite{nakahara2003,page1987,carlos2022}.

It is worth mentioning that the while deriving the expressions (\ref{metricoo}) and (\ref{twoformoo}) respectively for the metric components and K\"ahler form components we have assumed that the N-tuple $(w_1,w_2,\cdots,w_N)$ belongs to $\mathcal{N}^+$. It must be appreciated that any N-tuple in $\cc^N$ for which $\sum_{j=1}^{N} \l_{j} |w_{j}|^{2} > 0$ can always be mapped to a point in $\mathcal{N}^{+}$ by appropriate normalisation. So all N-tuples for which $\sum_{j=1}^{N} \l_{j} |w_{j}|^{2} > 0$ indeed have corresponding rays in the projective space $\mathscr{P}^{+}$, which contribute to the existence of the pseudo-K\"ahler manifold structure.

Similarly for any N-tuple in $\cc^N$ for which $\sum_{j=1}^{N} \l_{j} |w_{j}|^{2} < 0$ can always be mapped to set $\mathcal{N}^-$ by appropriate normalisation. One can define the projective space $\mathscr{P}^-$ by identifying the N-tuples in $\mathcal{N}^-$. One can immediately see that the projective space $\mathscr{P}^-$ will also display a pseudo-K\"ahler manifold structure like $\mathscr{P}^+$, with the same K\"ahler form and K\"ahler potential as (\ref{twoformoo}) albeit $\l_{j}$ being replaced with $-\l_{j}$.

One can also define the set of N-tuples for which $\sum_{l=1}^{N} \l_{l}|w^{l}|^{2} = 0$ as $\mathcal{N}^{0}$, and can have a corresponding projective space $\mathscr{P}^{0}$. These N-tuples correspond to the states in $\mathscr{K}_\oo$. It is evident that $\cc^N = \mathcal{N}^{+} \cup \mathcal{N}^{0} \cup \mathcal{N}^{-}$. It is worth noting that while the projective spaces $\mathscr{P}^{\pm}$ both are pseudo-K\"ahler manifolds, the projective space $\mathscr{P}^{0}$ is not a pseudo-K\"ahler manifold, since it does not admit the existence of metric $g^{\oo}$ and K\"ahler form $F^{\oo}$ owing to the fact that the term $\left( \sum_{l=1}^{N-1} \l_{l}|z^{l}|^{2} + \l_{N} \right) = 0$ in this space. It is interesting to note that while the metric and K\"ahler form are both non-singular in the projective space $\mathscr{P}^{\pm}$, the singularity can still be probed via points in $\mathscr{P}^{\pm}$ for which $\left( \sum_{l=1}^{N-1} \l_{l}|z^{l}|^{2} + \l_{N} \right) \ra 0$.

As an example let us consider the case of two state system, which was discussed earlier.  The observable $\oo$ of interest was taken such that its eigenvalues are $\l_1 = 1$ \& $\l_2 = -1$. A general state $\j$ can be represented by a pair of complex coefficients $(w_1,w_2)$ in the eigenbasis of $\oo$.  In the case, when $|w_1|^2 - |w_2|^2 > 0$, the image of $(w_1,w_2)$ lie in the projective space $\mathscr{P}^{+}$. All such pairs can be brought to $\mathcal{N}^+$, wherein  the complex coefficients are given by $w_1 = \pm \cosh \theta e^{i \phi}$ and $w_2 = \sinh \theta$. The inhomogeneous coordinate in such a case is $z = w_1/w_2 = \pm \coth \theta e^{i \phi}$. The  
corresponding K\"ahler potential reads $\mathscr{K}^{\oo} = \ln \left( |z|^2 - 1 \right)$, while the metric is given by:
\begin{align} \nmb
	g^{\oo} = \frac{-2 }{(|z|^{2} - 1)^{2}} d\bar{z} dz. 
\end{align}
When $|w_2|^2 - |w_1|^2 > 0$, the image of $(w_1,w_2)$ lie in the projective space $\mathscr{P}^{-}$, and all such pairs can be brought to $\mathcal{N}^-$. Here the complex coefficients are given by $w_1 = \sinh \theta e^{i \phi}$, $w_2 = \cosh \theta$; while the inhomogeneous coordinate $z' = w_1/w_2 = \tanh \theta e^{i \phi}$. One finds that the expressions for the K\"ahler potential and metric read the same as in the earlier case, albeit with the substitution $z \rightarrow z'$.

\section{Summary}

In this paper we study the mathematical structure underlying a quantum system that gives rise to the operator generalised geometric phase. The usual geometric phase is known to probe the symplectic structure of the ray space $\mathcal{R}$, which is formed by identifying unit normalised vectors, that differ by an overall global phase.  The ray space is known to possess a Riemannian metric, compatibly existing alongside the symplectic structure, making it a K\"ahler manifold. It is known that the ray space and the corresponding vector space together form a principal fibre bundle structure, and it is this structure that finds manifestation in the study of geometric phase. 

In this work we find that the operator generalised geometric phase does not capture the geometrical properties of the  space $\mathcal{R}$, or of the principal fibre bundle with $\ry$ as the base manifold. Rather the generalised geometric phase captures the 
geometry of the ray spaces $\ro^{\pm}$. The genesis of these ray spaces lies in working with a modified normalisation condition $(\j , \oo \j) = \text{sgn} (\j,\oo \j)$. This in general allows one to form two independent ray spaces $\ro^{\pm}$ by respectively identifying collinear vectors for which  $(\j , \oo \j) = \pm 1$. 
Corresponding to these ray spaces, the existence of two different principal fibre bundle structures with the base manifolds respectively as the two ray spaces $\ro^{\pm}$ is also found.  It is seen that akin to $\ry$, the ray spaces $\ro^{\pm}$ also admit a symplectic structure, as also  a pseudo-Riemannian metric, which is a generalisation of Fubini-Study metric.  This makes the ray spaces pseudo-K\"ahler manifolds. The generalised geometric phase $\gg_{\oo}$, as proposed in Ref. \onlinecite{vyas2023}, is now seen as (an)holonomy of connection $A^{\oo}$ over the fibre bundles. Digging further it is found that $\gg_{\oo}$ essentially captures the symplectic structure of the ray spaces $\ro^{\pm}$. It is evident that these ray spaces $\ro^{\pm}$ and the associated fibre bundles, go over to the usual case with ray space $\ry$ when $\oo \equiv \mathrm{1}$. 
Incidentally the geometry of pseudo-K\"ahler manifolds have attracted attention recently in the context of gravitation and cosmology \cite{carlos2022}.

This treatment essentially uncovers the geometry hidden in the average of an observable, which is fundamentally different from that of the manifold of pure state density matrices. This work shows that a quantum system can display a variety of different geometrical features depending upon the observable quantity that is being probed.

\section*{Acknowledgements} 
The author acknowledges beneficial conversations with Dr. M. Balodi and Prof. P. K. Panigrahi. 


%

\end{document}